\documentclass{article}
\usepackage{graphicx}
\usepackage{hyperref}

\input ibvs2.sty

\begin{document}
\IBVShead{5xxx}{7 June 2018}
	
\IBVStitle{TYC 5353-1137-1: an enigmatic Double Periodic Variable of}
\IBVStitle{semiregular amplitude}

\IBVSauth{ROSALES, J. A.$^1$, MENNICKENT, R. E.$^2$}
	
\IBVSinst{Astronomy Department, University of Concepci\'on, Concepci\'on, Chile. e-mail: jrosales@astro-udec.cl}\\

\IBVSinst{Astronomy Department, University of Concepci\'on, Concepci\'on, Chile. e-mail: rmennick@astro-udec.cl}\\	

\vskip 1cm
To date the Double Periodic Variables (DPVs) discovered by Mennickent et al. (2003) in the Large Magellanic Cloud (LMC) and the Small Magellanic Cloud (SMC) have been interpreted as semi-detached interacting binaries stars with a B-type component surrounded by an optically thick disk, these stars seem to experience regular cycles of mass loss (Mennickent et al. 2008) and are characterized by an orbital photometric variability on timescales of 2 to 100 days. These systems show a long period which is on average 33 times longer than the orbital period (Mennickent et al. (2016), Mennickent (2017), Poleski et al. (2010)). Currently, the DPVs found are Algol type eclipsing (DPV/E) and ellipsoidal (DPV/ELL) system. 

Therefore, we have performed a new search for DPVs of short period in the ASAS\footnote{http://www.astrouw.edu.pl/asas/} catalog (Pojmanski, G., 1997), focusing on those stars with orbital periods between 2 to 3 days which also show variations in their brightness. From a total of 244 objects, we have found another candidate to DPV, one whose mean brightness is gradually decreasing. By fitting a 3rd order polynomial to the mean magnitude (red line) and then moving it to zero for a second analysis, a gradual decrease over 2500 days was revealed. During the last 1000 days of this decrease, a 42\% increase in the variation between the minimum and maximum values of the magnitude was observed (Fig. 1). We determined the orbital period by using the PDM IRAF\footnote{IRAF is distributed by the National Optical Astronomy Observatories, which are operated by the Association of Universities for Research in Astronomy, Inc., under cooperative agreement with the National Science Foundation.} software (Stellingwerf 1978) and estimated the errors for the orbital period and long cycle by visual inspection of the light curves phased with trial periods near the minimum of the periodogram given by PDM. The two main frequencies of the system were disentangled using the code written by Zbigniew Ko\l{}aczkowski and described by Mennickent et al. (2012). This code was specially designed to adjust the orbital signal with a Fourier series and disentangle both frequencies using the fundamental frequencies and harmonics we supplied. The code removed this signal from the original time series thus allowing long periodicity to appear in a residual light curve, and we obtained both isolated light curves without additional frequencies, as shown in (Fig. 2, 3). We presented the search results and ephemeris in Table 1 and Fig. 1 (Left) both of which illustrate the gradual brightness decrease in the ASAS photometry. In the right panel of this figure we show the photometric variation $\Delta$V shifted to average zero and, finally, the disentangled light curves in Fig. 2 and 3. 

This enigmatic DPV presents a semiregular amplitude of the light curve when it is phased using the orbital period at three different photometric datasets (Fig 2.). The changes in the orbital light curve could be related to changes in disc size/temperature and spot temperature/position as proposed by Garc{\'e}s et al. (2018) for the DPV OGLE-LMC-DPV-097. Posteriorly, we disentangled the light curve of the long period and phased it. For that, we used the same time intervals as those used for the orbital period as a way to analyze possible variations in the amplitude of the light curve of this enigmatic phenomenon in the DPVs, and we apparently observed an effect of switch off-on of the long-cycle in the dataset of HJD between 2500 and 4000 (Fig. 3), this is observed for the first time in these type of systems. Therefore, we consider TYC 5353-1137-1 to be an optimal target for further photometric monitoring and spectroscopic studies, due to that will help us to test the mechanism based on cycles of the magnetic dynamo in the donor proposed by Schleicher \& Mennickent (2017), the cause of mass loss in some Algol stars and the evolutionary process of the DPVs.\\

\begin{center}
	\includegraphics[width=7.5cm,angle=0]{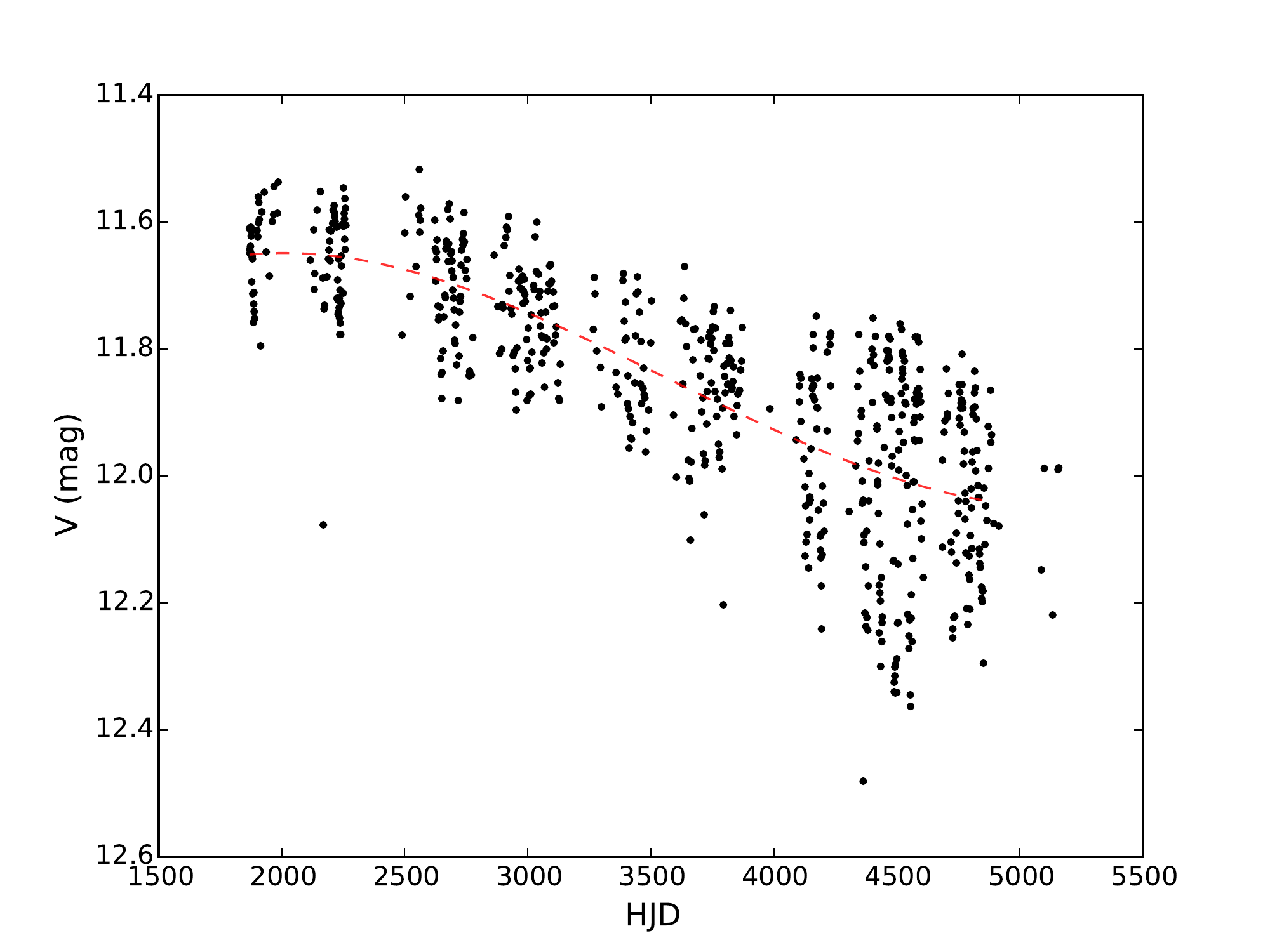}
	\includegraphics[width=7.5cm,angle=0]{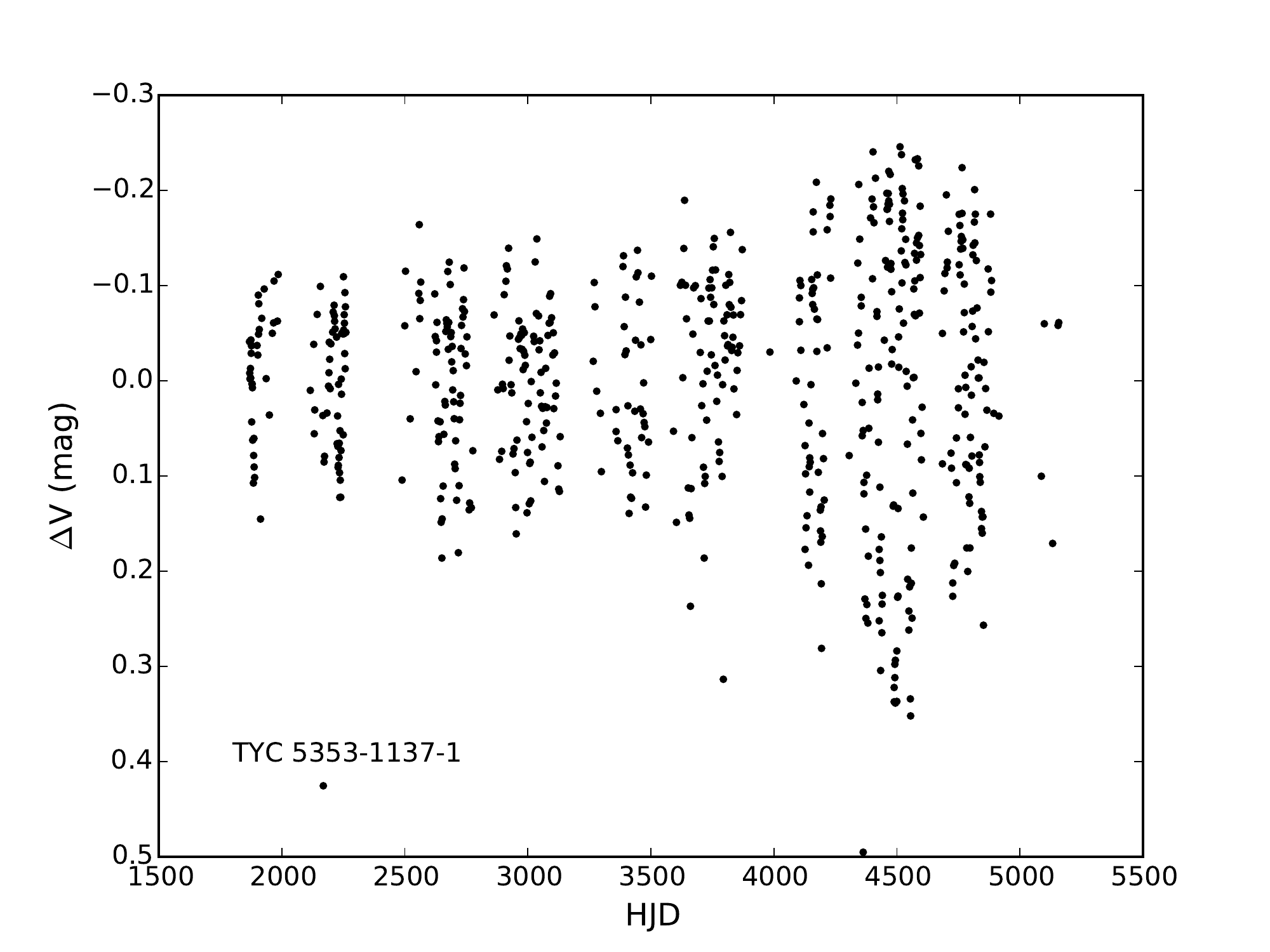}
\end{center}
{\bf Figure 1.}{(Left) The ASAS photometry reveals a gradual decrease of the DPV TYC 5353-1137-1 brightness during 2500 days, followed by an increase in the amplitude of photometric variation over the last 1000 days of 42\% (Right). The red line corresponds to a 3rd-order polynomial representing the mean magnitude.}

\vskip 0.5cm
\begin{center}
	\includegraphics[width=7.5cm,angle=0]{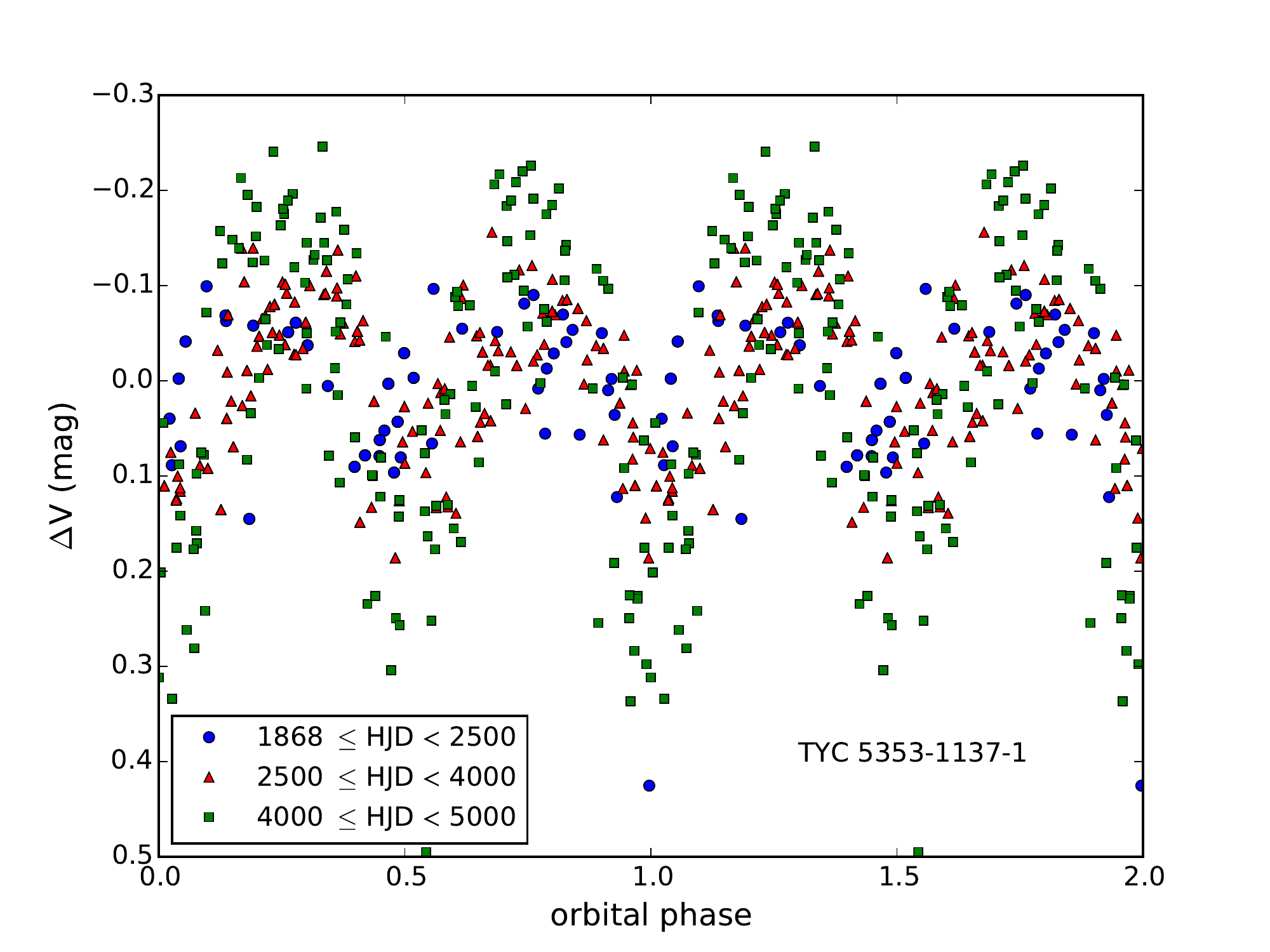}
	\includegraphics[width=7.5cm,angle=0]{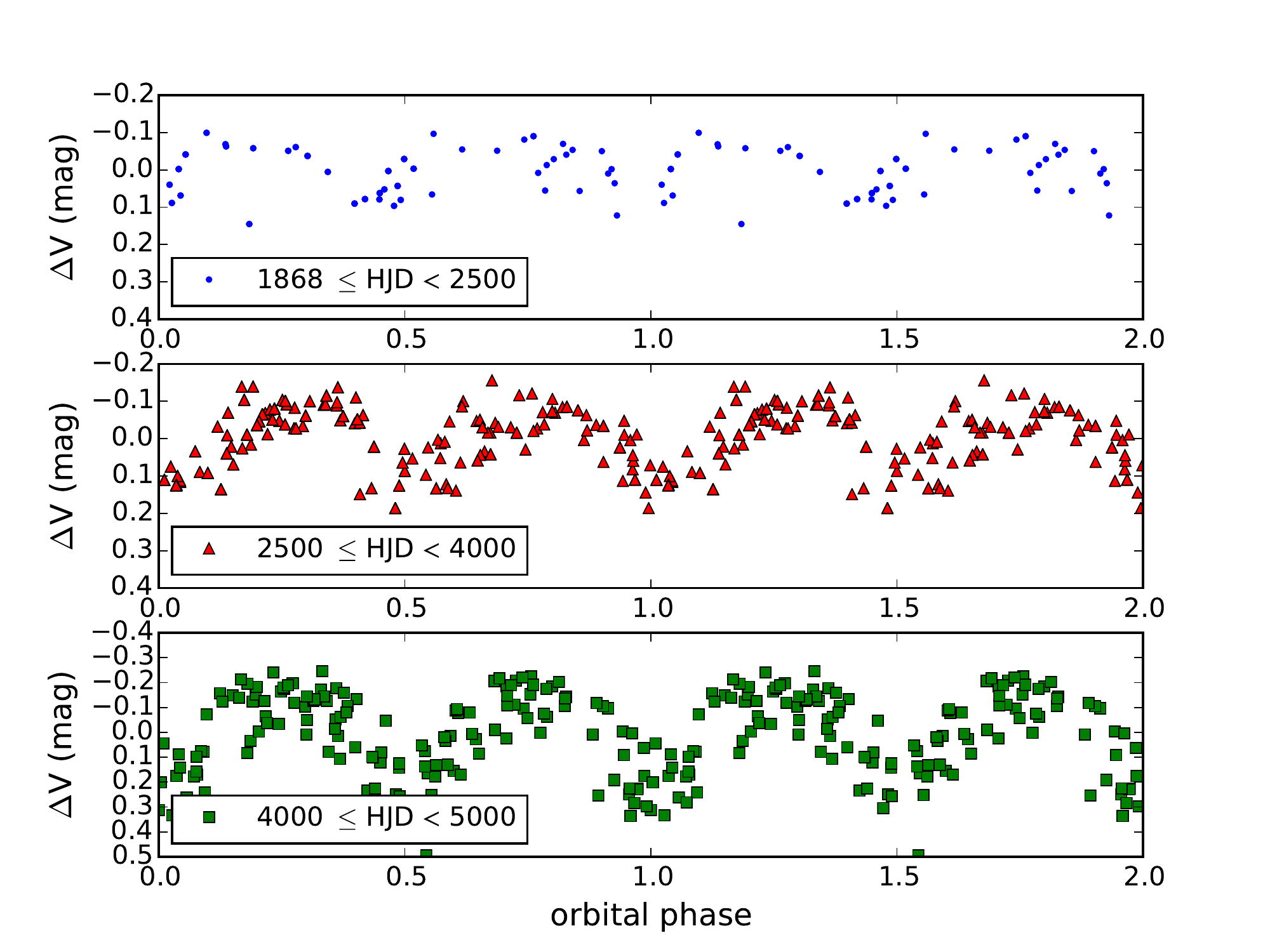}
\end{center}
{\bf Figure 2.} {Disentangled ASAS V-band light curve of the new confirmed Double Periodic Variable. The orbital phase has been separated in three datasets (-2450000.0), representing the variation of the amplitude.}

\vskip 0.5cm
\begin{center}
	\includegraphics[width=7.5cm,angle=0]{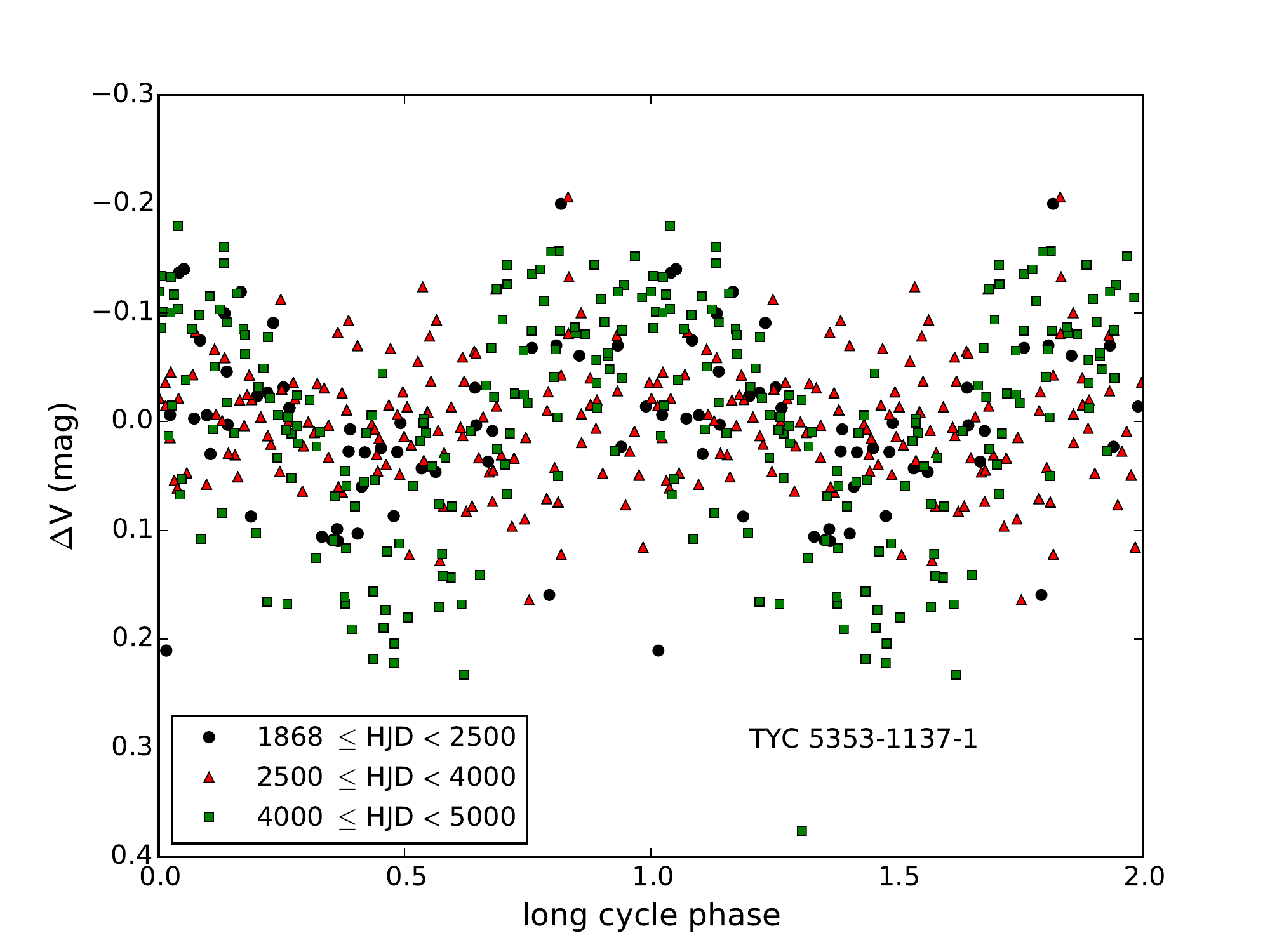}
	\includegraphics[width=7.5cm,angle=0]{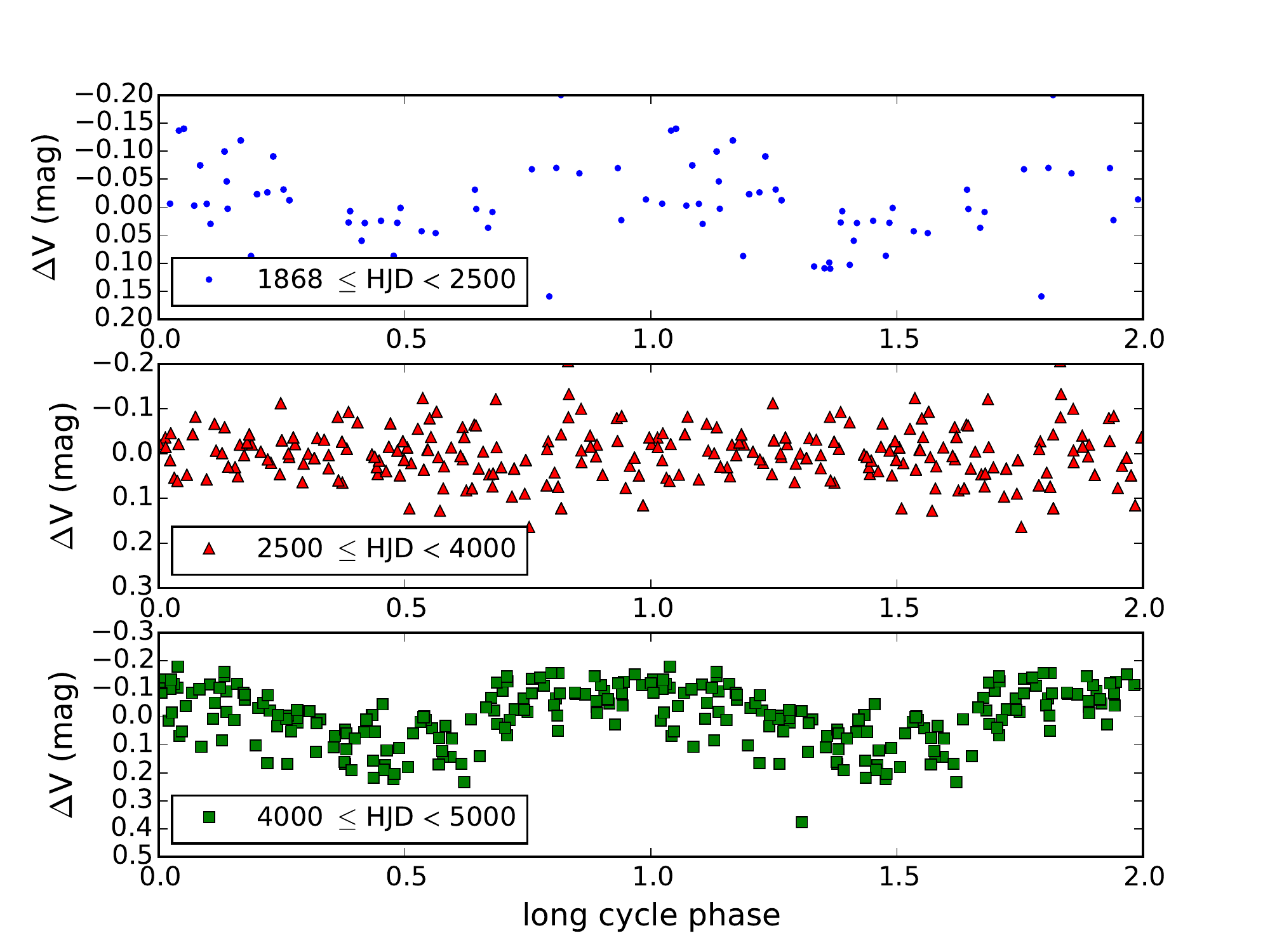}
\end{center}
{\bf Figure 3.} { The long cycle phase has been disentangled and separated in three datasets (-2450000.0). The first dataset shows less amplitude in the light curve of the long cycle (blue), during the second epoch occurs an effect of switch off (red), and the third dataset shows a remarkable increase in the amplitude of variability (green). Note the different y-axis scales in the panels.}

\vskip 0.5cm
\footnotesize
\begin{table}[ht]
	{\bf Table 1:} {Parameters of the newly confirmed DPV TYC 5353-1137-1 and its orbital ($P_{o}$) and long period ($P_{l}$). Epoch for both the minimum brightness of the orbital light curve and the maximum brightness of the long-cycle light curve are given.} \\ 
	\hspace{0.1cm} \\
	\scriptsize 
	\begin{center}
		\resizebox{16cm}{!}{$	
			\begin{tabular}{lcccccccc}
			\hline
			ASAS-ID		 & Other ID			& RA 			& DEC			& \textrm{P$_{o}$}	& \textrm{P$_{l}$}	& T$_{0}$(min$_{o}$)& T$_{0}$(max$_{l}$)& V (ASAS) 	\\
			{}			 & {}				& (2000)		& (2000)		& (days)	& (days)	& -2450000			& -2450000			& (mag)    	\\
			\hline
			{}			 & {}				& {}			& {}			& {}		& {}		& {}				& {}				& {}    	\\
			060418-1009.4& TYC 5353-1137-1	&  06:04:18.0	& -10:09:24.0 	& 2.028(1)	& 60.455(6)	& 4491.602390       & 4404.77653    	& 11.56	   	\\
			\hline 
			\end{tabular}
			$}
	\end{center}
\end{table}
\normalsize

\vskip 0.4cm

{\bf ACKNOWLEDGEMENTS}\\

\noindent
We acknowledge the anonymous referee whose comments helped to improve a first version of this report. R.E.M. gratefully acknowledges support by VRID-Enlace 218.016.004-1.0 and the Chilean Centro de Excelencia en Astrof{\'{i}}sica y Tecnolog{\'{i}}as Afines (CATA) BASAL grant AFB-170002. \\

\vskip 0.2cm
\references


{{Garc{\'e}s L}, J., Mennickent, R.~E., Djurasevi{\'c}, G.,Poleski, R., Soszy{\'n}ski, I., 2018, {\it MNRAS}, {\bf 477}, L11 \href{http://dx.doi.org/10.1093/mnrasl/sly042}{DOI}

Mennickent, R. E., Pietrzy{\'n}ski G., Diaz M., Gieren W., 2003, {\it A\&A}, {\bf 399}, L47 \href{http://dx.doi.org/10.1051/0004-6361:20030106}{DOI}

Mennickent, R. E., Ko{\l}aczkowski, Z., Michalska, G., et al. 2008, {\it MNRAS}, {\bf 389}, 1605 \href{http://dx.doi.org/10.1111/j.1365-2966.2008.13696.x}{DOI}

Mennickent, R.~E., Djura{\v s}evi{\'c}, G., Ko{\l}aczkowski, Z., \& Michalska, G. 2012, {\it MNRAS}, {\bf 421}, 862 \href{http://dx.doi.org/10.1111/j.1365-2966.2011.20363.x}{DOI}

Mennickent, R. E., Zharikov, S., Cabezas, M., et al. 2016b,  {\it MNRAS}, {\bf 461}, 1674 \href{http://dx.doi.org/10.1093/mnras/stw1416 }{DOI}

Mennickent, R.~E.\ 2017, {\it Serbian Astronomical Journal}, {\bf 194}, 1 \href{http://dx.doi.org/10.2298/SAJ1794001M}{DOI}

Pojmanski, G. 1997, {\it AcA}, {\bf 47}, 467 

Poleski R., Soszy{\'n}ski I., Udalski A., et al.  2010, {\it AcA}, {\bf 60}, 179 

Schleicher, D.~R.~G., \& Mennickent, R.~E.\ 2017, {\it A\&A}, {\bf 602}, A109 \href{http://dx.doi.org/10.1051/0004-6361/201628900}{DOI}

Stellingwerf, R. F. 1978, {\it ApJ}, {\bf 224}, 953 \href{http://dx.doi.org/10.1086/156444}{DOI}

\endreferences 

\end{document}